\documentclass[11pt,twoside]{article}
\usepackage{asp2004}
\usepackage{psfig}
\usepackage{epsf}
\usepackage{graphics}
\usepackage{lscape}
\def\edcomment#1{\iffalse\marginpar{\raggedright\sl#1\/}\else\relax\fi}
\reversemarginpar

\begin{document}
\title{The role of an accretion disk in AGN variability}
\author{B. Czerny}
\affil{Copernicus Astronomical Center, Bartycka 18, 00-716 Warsaw, Poland\\
       associated to Observatoire de  Paris, LUTH, 92195 Meudon, France}

\begin{abstract}
Optically thick accretion disks are considered to be important ingredients of luminous
AGN. The claim of their existence is well supported by observations and recent years
brought some progress in understanding of their dynamics. However, the role 
of accretion disks in optical/UV/X-ray variability of AGN is not quite clear.
Most probably, in short timescales the disk reprocesses the variable X-ray flux but
at longer timescales the variations of the disk structure lead directly to 
optical/UV variations as well as affect, or even create, the X-ray variability pattern.
We urgently need a considerable progress in time-dependent disk models to close the gap
between the theory and the stream of data coming from the AGN monitoring. 
\end{abstract}

\thispagestyle{plain}

\section{Introduction}

The broad band spectra of bright AGN clearly show the presence of both hot optically thin
plasma, responsible for hard X-ray emission, and the relatively cold plasma, 
presumably an optically 
thick accretion disk, dominating the optical/UV emission. The two components interact
radiatively, as proved by the presence of the so called X-ray reflection component
detected for the first time by Pounds et al. (1990). 

The exact geometry of the flow is still under discussion. The arguments for the presence of the 
cold accretion disk are circumstantial but rather strong. The geometry of the hot material
is less constrained and the dissipation processes within this plasma are poorly understood.

Strong variability observed in all spectral bands adds to the complexity of the accretion
process since the situation cannot be considered as basically stationary. On the other hand,
time dependence gives a direct insight into the dynamics of the flow, particularly into the
interaction between the hot plasma and the disk. 

\section{Accretion flow geometry}

Bright AGN show a Big Blue Bump (hereafter BBB) component that dominates the 
optical/UV emission. In Narrow Line Seyfert 1 galaxies and quasars this component
clearly extends to the soft X-ray band. BBB emission clearly comes from an optically
thick material, as seen from the presence of the Balmer edge (Kishimoto et al. 
2004). There are several arguments in favor of this component being roughly a Keplerian
disk: (i) in bright quasars and some NLS1 galaxies the predictions based on the simplest stationary
Keplerian black body disk reasonable represent this component (e.g. Koratkar \& Blaes 1999,
Soria \& Puchnarewicz 2002, Czerny et al. 2004); (ii) broad iron $K{\alpha}$ 
line in several NLS1 galaxies
is well described as originating from the matter in Keplerian motion (see Reynolds \& Novak 2003
for a review); (iii) $H{\beta}$ line in several
objects shows a disk-like component (see Eracleus, this proceedings) (iv) the spectral shape of BBB
in AGN is similar to the soft X-ray emission of many X-ray novae in their soft state where the
disk formation must take place since the mass supplied through the inner Lagrange point possesses
large angular momentum (v) jet formation indicates a disk-like geometry of the flow.

Power law character of the soft X-ray spectrum in NLS1 and quasars (Czerny \& Elvis 1987 and
subsequent papers), 
the change of the 
spectra slope at the Lyman edge position, and the absence of 
the Lyman edge itself in quasar spectra (Czerny \& Zbyszewska 1991, Blaes et al. 2001) 
are caused by Comptonization of the
disk radiation flux, either in the disk outer layers or in the surrounding hot plasma.
The disk in those objects extends most probably down to the marginally stable orbit. 

Such a disk seems to be absent in a very inactive nucleus like the center of our galaxy
(however, see Nayakshin, Cuadra \& Sunyaev 2003 for an opposite view), 
with its X-ray luminosity reaching now some $10^{36}$ 
erg s$^{-1}$ cm$^{-2}$ during periods of activity.

In intermediate luminosity objects, like radio-galaxies and normal Seyfert 1 galaxies,
the disk most likely exist in the outer part of the flow, at distances of few tens - few 
hundreds of Schwarzschild radii. The argument comes from interpretation of double profiles
of optical lines
(see Eracleus, this proceedings) and the presence of relatively narrow iron line (see
Reynolds \& Nowak 2003).
 
\begin{figure}
\begin{center}
\epsfxsize = 80 mm 
\epsfbox{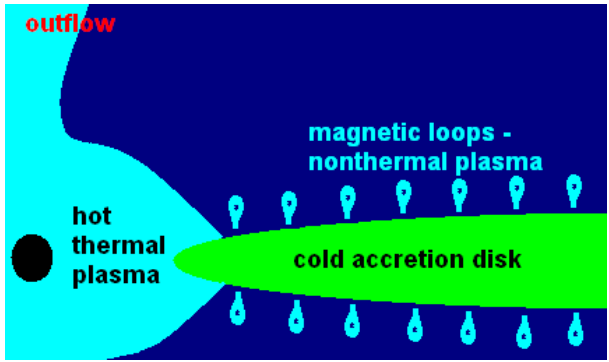}
\caption{The schematic view of an accretion flow: cold disk with magnetic flares above it in outer
region, replaced with the hot optically thin accreting plasma in the inner region, and 
accompanying outflow.
The transition radius is expected to decrease with increasing accretion rate.}
\label{fig:geom}
\end{center}
\end{figure}

Therefore, most broadly accepted view now is that the character of the flow is mostly
determined by the Eddington ratio. In high Eddington ratio objects accretion proceeds through a 
cold optically thick disk while in lower Eddington ratio the cold disk evaporates close to
a black hole, and below a certain radius, $r_{tr}$, depending on accretion rate, 
the accretion flow proceeds trough some form of optically thin hot flow. A plausible
geometry of the flow is shown in Fig.~\ref{fig:geom}. 

Even within this frame, several major open questions remain: (i) whether the Comptonizing plasma
responsible for formation of the soft X-ray tail of the BBB component is the also the source
of the hard X-ray emission; this is possible if plasma consists of mixture of thermal electrons
and non-thermal electrons (ii) whether hard X-ray emission actually mostly comes from the
magnetic coronal flares or from a standing shock at the basis of the outflow/jet.

Other flow geometries are also considered (for a review, see e.g. 
Collin et al. 2001), like quasi-spherical
inflow of optically thick bobs (Collin et al. 1996), cold disks extending always down to the marginally
stable orbit, with X-rays coming either from coronal flares (Galeev, Rosner \& Vaiana 1979) or 
from the base of a jet (Henri
\& Pelletier 1991). This last scenario predicts interesting effects due to the general relativity 
when the black hole is maximally
rotating and the jet base is very close to the black hole (Miniutti \& Fabian 2004). 
Without precise modeling, including variability studies, it
is not possible to differentiate between these alternatives.

\section{Basic local timescales of Keplerian flow}
\label{sect:timescales}

In this section I will summarize the basic timescales characteristic for a Keplerian,
geometrically thin and optically thick disk as a function of the distance from the
black hole and other parameters. Some of these timescales are considerably model-dependent,
as notified. 

\subsection{Dynamical timescale}

If the accretion flow is roughly Keplerian, the dynamical timescale of the matter is given
by the Keplerian frequency:
\begin{equation}
t_{dyn}=\Omega_K^{-1}={\sqrt {GM \over r^3}}
\end{equation}
determined by the mass of the black hole, $M$, and the radius $r$. This timescale, equal to 
the orbital period, describes the motion at the circular orbit, the local rotation with the 
epicyclic frequency (for example, if a magnetic loop emerges from the disk surface, its foots 
entangle in this timescale), the dynamical oscillations in the direction perpendicular to the 
disk surface, the timescale to achieve the hydrostatic equilibrium in the disk and the
sound crossing timescale in the disk, in the vertical direction. The free fall timescale
from the radius $r$ towards the black hole is also of the same order of magnitude.

Very close to a black hole, the Keplerian frequency, the epicyclic frequency and the oscillations
in the vertical direction differs due to the effects of General Relativity (see e.g. Kato 2001). At
distances larger than 10 $R_{Schw}$ we can neglect these effects.

We can conveniently express the dynamical timescale using $R_3 = r/ (3 R_{Schw})$ as dimensionless
units of radius and $M_8 = M/(10^8 M_{\odot})$ as dimensionless units of mass
\begin{equation}
t_{dyn}=10^4 R_3^{3/2}M_8~~[{\rm s}]. 
\end{equation}

The timescale of propagation of the sound waves in the radial direction, $t_{sound-r}$ is longer
\begin{equation}
t_{sound-r} = t_{dyn} \bigl({r \over h_d}\bigr),
\end{equation}
where $h_d$ is the disk thickness.

\subsection{Thermal timescale}

Thermal timescale of the disk can be determined if the heating and cooling mechanisms are 
specified since this timescale is defined as a ratio of internal energy to the cooling or
heating rate. Here we adopt the assumption that the disk viscosity is described by the parameter
$\alpha$ introduced by Shakura \& Sunyaev (1973). The support for this idea is discussed in more
detail in Sect.~\ref{sect:MRI}. If the assumption that the stress tensor is equal to $\alpha P$
is introduced, where $P$ is the total pressure, the characteristic thermal timescale of the disk
is given by
\begin{equation}
t_{th}=\alpha^{-1} t_{dyn}.
\end{equation}
This timescale does not depend on the optical depth of the disk or the cooling mechanism so we
have the same thermal timescale for a cold optically thick disk and a hot optically thin flow.
Assuming $\alpha = 0.1$ as characteristic value of the viscosity parameter, we obtain
\begin{equation}
t_{th}=10^5 \alpha_{0.1}^{-1}R_3^{3/2} M_8~~[{\rm s}].
\end{equation}

\subsection{Viscous timescale}

Viscous timescale is defined as a characteristic timescale of mass flow, i.e. locally as 
the ratio of the radius to the radial velocity. If we assume the $\alpha$ disk model, we can
obtain the following, more convenient expression 
\begin{equation}
t_{visc}= t_{th}\bigl({r \over h_d}\bigr)^2.
\end{equation}
For a cold optically thick disk, the $h_d/r$ ratio is small
so the viscous timescale is orders of magnitude longer than the thermal timescale. Even if the
Eddington ratio of an object is close to 1, this ratio remains relatively small above $\sim
10 R_{Schw}$. However, for highly super-Eddington flow  or for a hot optically thin plasma
at virial temperature $h_d/r$ ratio is close to 1 and the viscous timescale of such a flow
is equal to the thermal timescale. 
Adopting $h_d/r=0.1$ as a characteristic value, we can write
\begin{equation}
t_{visc}=10^7 \alpha_{0.1}^{-1}\bigl({r \over 10 h_d}\bigr)^2 R_3^{3/2}M_8~~ [{\rm s}].
\end{equation}

\begin{figure}
\begin{center}
\epsfxsize = 80 mm 
\epsfbox{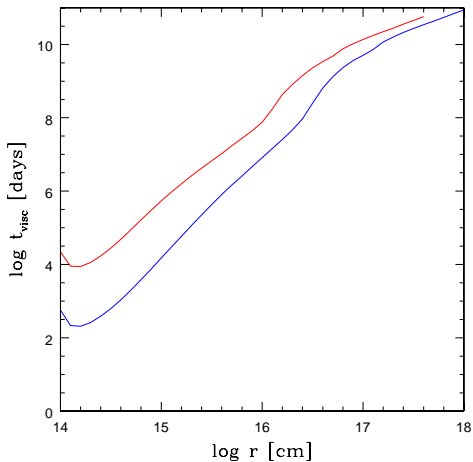}
\caption{Local viscous timescale of an accretion disk around $10^8 M_{\odot}$ black hole for
the Eddington ratio 0.3 (lower curve) and 0.03 (upper curve), and the viscosity parameter
$\alpha = 0.02$}.
\label{fig:visc}
\end{center}
\end{figure}

Actually, the disk thickness depends is roughly given by $h_d = 10 \dot m$, in the inner, radiation
pressure dominated region, and depends both on the accretion rate and the disk radius further out.
Estimates are influenced by the description of the disk opacity. Examples of numerical results are
shown in Fig.~\ref{fig:visc}, computed using the disk structure code of R\' o\. za\' nska et al. (1999). 

\subsection{Timescales of X-ray reprocessing}

Variations in the X-ray flux, generated either in the magnetic loops above the disk or in the
innermost part of the flow, leads to changes of the condition in the disk surface layers.
The timescales of the processes involved can be estimate as functions of the local number density, 
$n$,temperature, $T$,
disk radiation flux, $F_{soft}$ and the cooling function, $\Lambda$, in the disk surface layers, 
as discussed by Collin et al. (2003). Introducing the dimensionless parameters we obtain the
following estimates.
The characteristic timescale for the radiation transfer is
\begin{equation}
t_{rt}={\tau_{es} (\tau_{es}+1)\over \sigma_T c n} \approx 100 n_{12}^{-1}~~[{\rm s}],
\end{equation}
the ionization state of the disc surface adjusts on a timescale 
\begin{equation}
t_{ion}={h \nu \over F_X \sigma_{ion}} \approx 10^{-7} F_{16} ~~[{\rm s}]
\end{equation}
and recombination proceeds in the timescale
\begin{equation}
t_{rec}={1 \over n \alpha_{rec}} \approx 10n_{12}^{-1}~~[{\rm s}].
\end{equation}
The thermal timescale of the disk surface layers depends on the ionization state and, consequently,
on the cooling mechanism. If the ionization is weak or moderate, atomic cooling dominates and the
timescale of restoring the thermal equilibrium is given by
\begin{equation}
t_{th-surface}^{atomic} = {nkT \over n^2 \Lambda} \approx 10T_6n_{12}^{-1}\Lambda_{23}^{-1}~~[{\rm s}]
\end{equation}
while in the case of very strong ionization the Compton cooling dominates (the temperature of the 
disk surface is in this case close to the Compton temperature) and the timescale is
\begin{equation}
t_{th-surface}^{Compton}={nkT \over F_{soft}{4kT \over m_ec^2} \sigma_T n} \approx 10^4 F_{14}^{-1}~~[{\rm s}].
\end{equation}
Heating of the disk surface affects the hydrostatic equilibrium and the characteristic timescale for
expansion or contraction of the disk surface layers is given by
\begin{equation}
t_{dyn-surface}={H \over c_s} \approx 10^5 T_6^{-1/2}n_{12}^{-1}~~[{\rm s}].
\end{equation}
We see that these timescales are generally very short, apart from the last one which may be 
comparable, or longer than the dynamical timescale of the disk body.

\subsection{Timescale of cold disk removal}

If accretion flow proceeds as shown in Fig.~\ref{fig:geom}, and the transition radius changes with
time, we need an estimate of the characteristic timescale of the removal of the cold disk from
a given radius down. This removal can happen either in a form of evaporation and a change of
accretion flow into an optically thin flow, or in a form of ejection (outflow). In both cases we
require the change of the disk temperature to roughly the virial temperature and we need to
accumulate enough energy from the accretion flow for the transition to happen. Assuming the
$\alpha$ viscosity disk model, we obtain
\begin{equation}
t_{evap} = {E \over F} = {{k \over m_H} \Sigma T_v \over \alpha H P \Omega_K } \equiv \tau_{visc},
\end{equation}
so the timescale is the same at the cold disk viscous timescale at a given radius. We can also obtain 
more general estimate, without assuming $\alpha$ disk. 
\begin{equation}
t_{evap} = {E \over \eta \dot M c^2}; ~~~~E = \pi r^2 \Sigma {k \over m_H} T_v; ~~~~\eta = {r \over 4 R_{Schw}},
\end{equation}
which can be expressed conveniently as
\begin{equation}
t_{evap} = 1000 \bigl({r \over 100 R_{Schw}} \bigr)^2\dot m_{0.1}M_8~~[yr].
\end{equation}
The process is therefore long; observations show that in galactic sources state
transitions, believed to be due to the cold disk removal, last one day, 
and in AGN they should take thousands of years.

Therefore, if the inner disk seems to disappear in a timescale of months -- years, we
should rather interpret it as temporary suppression of the disk emission, either due
to the excessive cooling or due to suppression of the energy dissipation, e.g. 
turning off the MRI instability by temporary formation of strong ordered magnetic field
(see Marsher, this proceedings).

\section{Basic non-local aspects}

The local state of the plasma, at any radius, is affected by the processes taking place
elsewhere. The influence propagates both in $\rightarrow$ out and out $\rightarrow$ in. 

The first
class includes (i) irradiation of the outer disk by the radiation generated in the inner
part (either direct or indirect, through the scattering of some part of emission in the
optically thin plasma present in the inner region) (ii) mechanical transport of energy in the form
of convection in the optically thin flow (e.g. CDAF models of Narayan, Igumenshchev \&
Abramowicz 2000), or wind/coronal outflow with large angular momentum (e.g. Cao \& Spruit 1994; 
Janiuk \& Czerny 2004). Irradiation, or any kind of additional energy transport,
significantly complicates the use of the timescales discussed in Sect.~\ref{sect:timescales}.
In the case of strongly irradiated disk, the emission at a given wavelength comes predominantly
from much larger radius than predicted by the model without irradiation, and the observed
variability may contain both the contribution from the variable irradiation and the intrinsic
variability of the disk.

The second class includes (i) modulations of accretion rate (ii) coronal inflow. 
The first effect may be caused by external perturbations of the flow but
certain class of modulations is predicted to develop internally (see Sect.~\ref{sect:MRI}.
The second effect may be present if significant angular momentum transfer
can take place in the disk corona itself. Since the viscous timescale in a
hot corona may be comparable to its thermal timescale such surface inflow may
proceed much faster than the inflow through the disk main body. 

\section{Theory of disk instabilities}

Disk instabilities do not necessarily destroy the accretion disk, as sometimes
believed. In opposite, they seem to provide the explanation for some aspects
of the accretion disk existence and behaviour.

\subsection{Magnetorotational instability}
\label{sect:MRI}

This instability is now established as the physical mechanism of the 
accretion disk viscosity (Balbus \& Hawley 1991). Advanced MHD computations
show that this instability roughly corresponds to $\alpha \sim 0.1$ in
gas pressure dominated flow (Hawley \& Krolik 2001). However, besides providing
the time and spatially averaged effective viscosity, the instability produces
(i) local fluctuations in dissipation and the accretion rate 
(ii) certain level of disk clumpiness (iii) specific
vertical stratification of the dissipation. 

The first effect is now 
considered as an attractive qualitative explanation of the power law type
shape of the power spectra in X-ray energy band in objects like galactic
sources in soft state and AGN with the same shape of power spectra 
(Lyubarskij 1997, King et al. 2004). 
The second effect may provide the explanation for the apparent variable 
clumpiness of the disk
requested to explain the variations of the emission line profiles (Asatrian, 
this proceedings;
Eracleus, this proceedings; Sergeev, this proceedings, Shapovalova, 
this proceedings).
The third
effect leads in a natural way to formation of either strong magnetic flares
above the disk body, 
or magnetically heated upper disk skin of moderate optical depth. The first
result was obtained when the cooling was neglected in MHD simulations (Miller
\& Stone 2000) while the second one was obtained for a radiation pressure dominated
medium with flux-limited approximation for cooling (Turner 2004). The 
issue will be resolved when MHD simulations are performed with Compton cooling
included.  

MHD simulations of Miller \& Stone also show that large loop formation 
takes more than one dynamical timescale so the timescale of the corona
formation may be between the dynamical timescale and the thermal timescale 
of the disk body.

\subsection{Radiation pressure instability}

Standard $\alpha$ disk models are unstable if dominated by radiation pressure 
(Pringle, Rees \& Pacholczyk 1973). AGN disk models show the domination 
by radiation pressure for a broad range of Eddington rates. Disk evaporation
is not likely to prevent the existence of such a disk region (e.g. R\' o\. za\' nska
\& Czerny 2000). Computations of disk time evolution under the influence of such
instability show semi-regular outbursts in viscous timescale of the inner 
$\sim 100 R_{Schw}$ (Szuszkiewicz \& Miller 1998, Teresi, Molteni \& Toscano 2004; 
see also Janiuk, Czerny \& Siemiginowska 2000 
in the context of GRS 1915+105). Such effect is seen only in one galactic
source, GRS 1915+105, which most probably has the highest Eddington rate
(Done, Wardzi\' nski \& Gierli\' nski 2004), but
neither in other galactic sources nor in AGN. Scaling GRS 1915+105 outbursts,
lasting 100 - 1000 s to black hole mass $10^7 M_{\odot}$ we would expect 
outbursts lasting 3 yr - 30 yr. 

The redistribution of the dissipation
in the vertical direction due to MRI instability and magnetic energy transport
may suppress this instability (e.g. Czerny et al. 2003). MHD simulations
by Turner (2004) mentioned in Sect.~\ref{sect:MRI} indicate that the instability
is not completely dumped out but partially suppressed: instead of large
outbursts we should expect rather irregular variability by a factor of a few, roughly
in thermal timescale.    

\subsection{Ionization instability}
\label{sect:ion}

Present in outer parts of the disk in galactic sources (X-ray novae and dwarf novae; 
for a review, see Lasota 2001). 
It remains an open question whether the mechanism applies to AGN. In any case, expected
timescale are thousands to millions of years (e.g. Janiuk et al. 2004). 

\section{Time-dependent disk models}

Most of the work was devoted so far to the stationary disk models and to proper modelling
of their spectra. A few existing models were aimed at studying some specific effects.

Global evolution of accretion disks under the radiation pressure instability was studied by 
Honma, Matsumoto \& Kato (1991), Szuszkiewicz
\& Miller (1998), Nayakshin, Rappaport \& Melia (2000), Szuszkiewicz \& Miller (2001), 
Janiuk et al. (2002), Teresi, Molteni \& Toscano (2004).

Global evolution of AGN disks under ionization instability was calculated by 
Mineshige \& Shields (1990), Siemiginowska, Czerny \& Kostyunin (1996), Hatziminaoglou, Siemiginowska
\& Elvis (2001), Janiuk et al. (2004), and a lot of work was done in this field for binary stars,
as mentioned in Sect.~\ref{sect:ion}.

Pringle (1997) calculated the evolution of the disk under irradiation-induced warp instability.
Hujeirat \& Camenzind (2000) calculated the cold disk disruption and formation of a hot inner torus.

An interesting cellular-automaton model was developed by Mineshige, Ouchi \& Nishimori (1994),
and applied subsequently to analyze the properties of the optical AGN lightcurve (e.g. Kawaguchi et al. 
1998; see Hawkins, this proceedings).

\section{The origin of observed optical/UV variability}

Practically all radio quiet AGN show some variability in the optical/UV band.
This variability may be due to (i) variable X-ray irradiation, (ii) intrinsic disk variability
caused by disk instabilities, (iii) variable obscuration. It is most probable that
more than one mechanism is in action.  An idea of separate slow and fast variability 
was advertised by Lyutyi (this proceedings) and de Vries (this proceedings). Similarly,
the comparison of the power spectra in the optical and X-ray band indicate that at shorter
timescales the optical variations are caused by irradiation but at longer timescales 
of years there seems to be an excess of the optical variations (Czerny et al. 1999, 
Czerny et al. 2003b).

\subsection{variable X-ray irradiation}

Results of many monitoring projects, with good coverage of short timescale variability, 
have been discussed during this conference.
Observations of Seyfert galaxies in the optical band show relative delays of
continuum emission in various wavelengths by a day or a few days. The results
are roughly consistent with variable X-ray irradiation (Sergeev, this proceedings). 
Similar conclusion
was reached for IR monitoring, and the delays with respect to the optical
emission were naturally longer (Oknyanskij, this proceedings). In this case the role of the
disk is mostly passive in this sense that character of the variability is determined
by the intrinsic timescales of the X-ray emitting plasma plus the light travel time.

Quasars are also variable in the optical band when monitored for several years 
(see Papadakis \& Magotis; Hawkins; de Vries, this proceedings). Delays were not determined so far so 
we have less direct constraints of the variability mechanism. In quasars the X-ray emission is 
relatively less important than in Seyfert 1 galaxies
so we might expect relatively larger role of the intrinsic disk variability.

As for the theory of disk irradiation, there has been a considerable progress in modeling of the 
X-ray reprocessing, including
the stratification of the disk surface layers (e.g. Goosmann, this proceedings). 
New models start to pay attention to the time-dependent response of the disk surface layers (e.g.
Nayakshin \& Kazanas 2002, Collin et al. 2003, Czerny \& Goosmann 2004). Also the angular-dependent
emissivity of the hot material should be studied. Poor correlation between primary X-ray emission and
X-ray reflection is frequently seen in the data. For example,  
optical events without an X-ray counterpart were
observed in Akn 564 which may mean that the hot plasma emission may be strongly anisotropic 
(see Gaskell, this proceedings).

\subsection{intrinsic variations in disk dissipation}

However, when a combined X-ray and UV monitoring is performed, a more complicated
picture emerges (Uttley, this proceedings; Arevalo, this proceedings). 
We certainly see some reprocessing in shorter
timescales but at somewhat longer timescales we have an additional effect of optical emission 
{\it  leading} X-ray variability. Not many such examples are known because of the difficulties
in the proper data coverage. 
The optical variations leading by a few days detected in 
NGC 4051 ($M = 5 \times 10^5 M_{\odot}$) would correspond to timescales of a few hundred days 
for NGC 5548 or NGC 4151. The presence of such delays indicate that changes in the disk
interior lead to changes in the disk emission and subsequent changes in the state of the hot
material. The candidates for the prime cause of those variations are either MRI turbulence, 
or strongly suppressed radiation pressure instability, as in simulations of Turner (2004).
The local effective timescale is in this case not far from the thermal timescale. This possibility
was explored by Starling et al. (2004). They interpreted the optical/UV variability in a sample of 
PG quasars
as happening in local thermal timescale and determined the value of the viscosity parameter.
Obtained value, $\alpha \sim 0.02$ is reasonable. However, other interpretations of quasar variability
are still open.

\subsection{Intrinsic variability vs. variable obscuration as the cause of large variations}

Many AGN show occasional dramatic changes in their properties, including
the change in their classification (e.g. from LINER to Seyfert, Yuan et al. 2004; Seyfert
class change, e.g. Lyutyi, this proceedings; switch off the X-ray source, e.g. Guainazzi et al. 
1998).
Monitoring brings more and more of such examples. Therefore, one of the
big questions discussed during this conference was: are these changes
intrinsic or do they result from variable obscuration?

The success of the reverberation approach to the analysis of the BLR (Peterson,
this proceedings) shows that strong intrinsic variability is certainly present.
Also hints for asymmetry of the variability (rise time frequently shorter that the
decay time; see Lyutyi, this proceedings; Hawkins, this proceedings) supports the
intrinsic character of the luminosity variations. However, it does not necessarily mean
that variable obscuration is absent.  

Statistical studies show that highly obscured AGN are four times more
numerous than unobscured (e.g. Treister et al. 2004). Obscuration is most
probably very important also in apparently unobscured objects like
NLS1 (e.g. Constantin and Shields 2003) or quasars (e.g. Czerny \& Li 2004). 
Variable obscuration in single objects was claimed
to be seen in a number of sources (Risaliti et al. 2002), and partial covering
was successfully applied e.g. to NLS1 IRAS 13224-3809 by Boller et al. (2003).
Also interestingly, 9 year observations of NGC 4151 with BATSE in $\gamma$-ray band
show variability (Hill et al. 2004) but does not seem to follow the 10-year timescale of outburst
seen in this source between 1991 and 2000. It might mean that slow variability, 
strongly seen in optical/UV and less in X-ray band is mostly due to variable 
extinction. Strongly variable extinction is also an important feature of a class
of galactic sources known as dippers (e.g. 'Big Dipper'; Smale et al. 2001). 

Therefore, we possibly have a variable extinction as well, perhaps resulting from 
the changes in gas ionization and dust evaporation or sublimation.

\section{Summary}

Cold, geometrically thin, optically thick accretion disk certainly plays a role in the
AGN variability. It reprocesses the variable X-ray emission of the coexisting hot, 
optically thin plasma in a complex way, thus playing mostly a passive role in the variability at the 
shortest timescales. There are arguments, however, supporting the idea of an active role
of the disk at longer timescales, and the internal disk variability may be possibly 
connected with the MRI instability and/or partially saturated radiation pressure instability.
These changes, in turn, may propagate within the disk and affect the hot X-ray emitting plasma. 
Determination of the nature of the optical/UV variations need not only further monitoring, but
also a development of the time-dependent disk models, including the non-local phenomena.
The description of X-ray reprocessing is already quite advanced but the reasonable
description of time-dependent flow is still missing. MHD simulations have poor description of cooling
and cannot be extended to timescales comparable to the viscous timescale for technical reasons,
while simpler models based on $\alpha$ prescription still miss important ingredients like
better approximation of the vertical stratification of the dissipation and the role of the magnetic
field in energy transport.

\acknowledgements

Part of this work was supported by grant 2P03D00322 of the Polish
State Committee for Scientific Research and by 
and by Laboratoire Europe\' en Associ\' e Astrophysique Pologne-France.

\end{document}